\documentclass[aps,twocolumn,english,prl]{revtex4-1}

\usepackage[T1]{fontenc}
\usepackage[latin9]{inputenc}
\usepackage{textcomp}
\usepackage{amsmath}
\usepackage{graphicx}
\usepackage{amssymb}
\usepackage{babel}

\begin{document}

\title{Strain-mediated coupling in a quantum dot-mechanical oscillator hybrid system}

%
%

\author{I. Yeo$^{1}$, P.-L. de Assis $^{1}$, A. Gloppe $^{3}$,
E. Dupont-Ferrier$^{3}$, P. Verlot$^{3}$, N.S. Malik$^{2}$, E. Dupuy$^{2}$, J. Claudon$^{2}$, J.-M. G\'{e}rard$^{2}$, A. Auff\`{e}ves$^{1}$, G. Nogues$^{1}$, S. Seidelin$^{3}$, J.-Ph. Poizat$^{1,*}$, O. Arcizet$^{3,*}$, and M. Richard$^{1}$}

\affiliation{ $^1$ 'Nanophysics et Semiconductors' joint team, Institut N\'{e}el, CNRS -
Universit\'{e} Joseph Fourier, 38042 Grenoble, France, \\
$^2$ 'Nanophysics et Semiconductors' joint team, CEA/INAC/SP2M, 38054 Grenoble, France, \\
$^3$ Institut N\'{e}el, CNRS and Universit\'{e} Joseph Fourier, 38042 Grenoble, France,\\
* Corresponding authors : olivier.arcizet@grenoble.cnrs.fr, jean-philippe.poizat@grenoble.cnrs.fr}


\maketitle

\textbf{Recent progress in nanotechnology has allowed to fabricate new hybrid systems where a single two-level system is coupled to a mechanical nanoresonator \cite{Treutlein,Lahaye,Hammerer,Rabl,Hunger,Bennett}. In such systems the quantum nature of a macroscopic degree of freedom can be revealed and manipulated. This opens up appealing perspectives for quantum information technologies \cite{Palomaki}, and for the exploration of quantum-classical boundary. Here we present the experimental realization of a monolithic solid-state hybrid system governed by material strain \cite{Wilson-Rae}~: a quantum dot is embedded within a nanowire featuring discrete mechanical resonances corresponding to flexural vibration modes. Mechanical vibrations result in a time-varying strain field that modulates the quantum dot transition energy. This approach simultaneously offers a large light extraction efficiency \cite{Claudon,Claudon2010} and a large exciton-phonon coupling strength $g_0$. By means of optical and mechanical spectroscopy, we find that $g_0/2\pi$ is nearly as large as the mechanical frequency, a criterion which defines the ultra-strong coupling regime \cite{Armour}.}


\begin{figure}[t!]
\includegraphics[width=0.9\columnwidth]{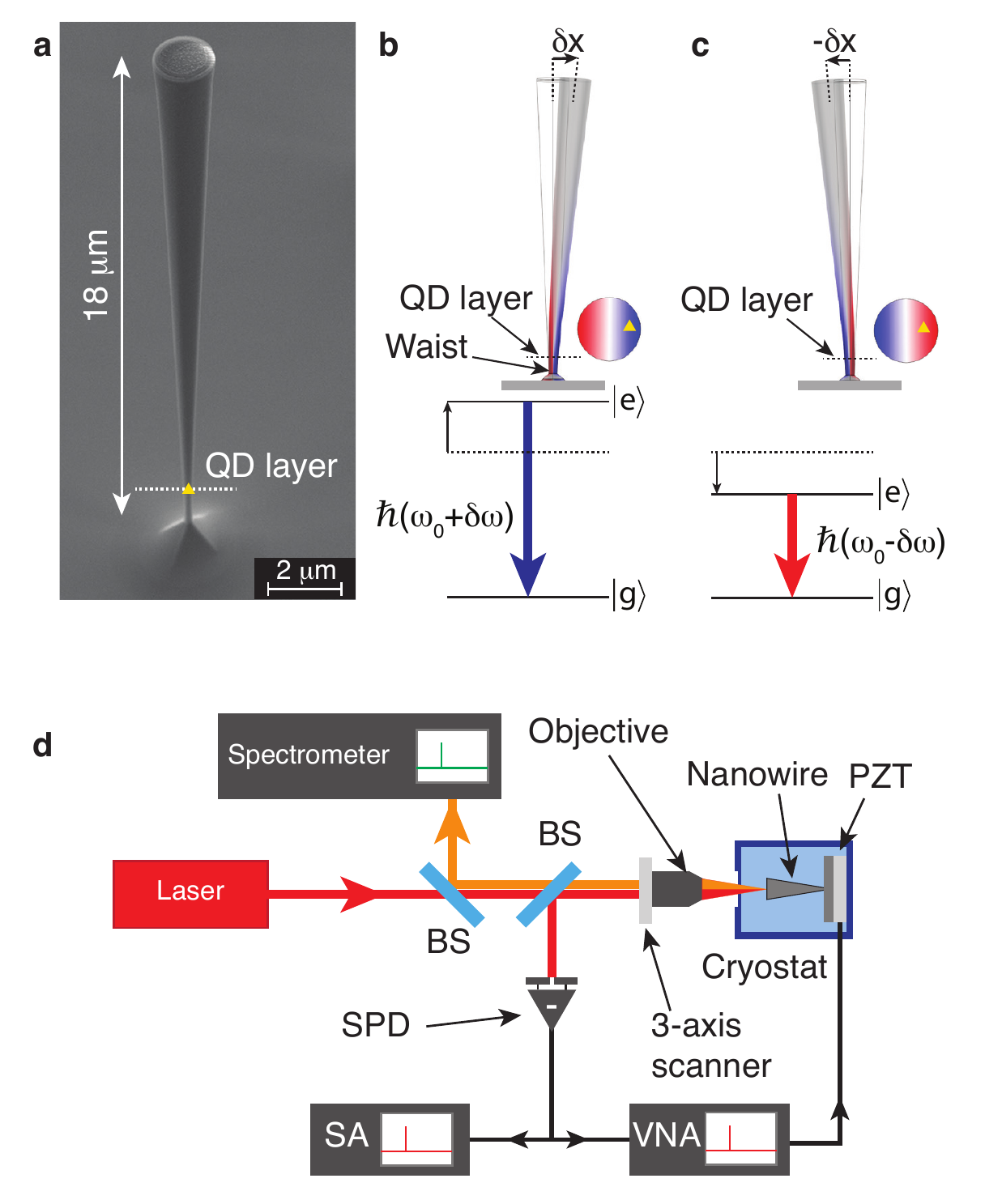}
\caption{\textbf{Hybrid system and experimental setup.} \textbf{a} Scanning electron microscope picture of a representative cone shaped nanowire. The quantum dots (QDs) layer is materialized by the dashed white line. \textbf{b} and \textbf{c}, Nanowire deformation in the first order flexural vibration mode. The stress field is plotted in blue to red color scale: due to its excentric in-plane position, the quantum dot (yellow triangle) experiences in \textbf{b} a compressive strain that shifts its transition energy $\hbar\omega_0$ by $+\hbar\delta\omega$ and in \textbf{c} a tensile strain that shifts its transition energy by $-\hbar\delta\omega$. \textbf{d} Experimental setup: single QD optical measurements are carried out using a spectrometer, and the measurement of the nanowire free-end displacement $\delta x$ is realized by means of a balanced split photo-diode (SPD). The voltage difference $v$ between the quadrants is sent either to a vectorial network analyzer (VNA) or to a spectrum analyzer (SA).}
\label{fig1}
\end{figure}

A single quantum two-level system coupled to a micron-size mechanical oscillator constitutes a hybrid system, which connects two different worlds: the classical and the quantum one. This new kind of interaction opens up the possibility of creating macroscopic non-classical states of motion, such as phonon Schr\"{o}dinger cats or phonon number states. In the case of strain-mediated coupling, it is predicted that the two level system can even be used to cool the mechanical resonator down to its ground state \cite{Wilson-Rae} or conversely to achieve phonon lasing \cite{Kabuss}.

Such appealing prospects have recently motivated the development of several kinds of hybrid systems, like for instance: (i) a single spin embedded in a mechanical resonator coupled together by an external magnetic field gradient \cite{Rabl,Arcizet,Kolkowitz}, (ii) a few elementary charges (single electron or Cooper pair) coupled by electrostatic forces with a vibrating gate \cite{Lahaye,OConnell,Bennett}, or (iii) quantized current loops in superconducting qubits attached to a mechanical oscillator interacting via a magnetic field \cite{Etaki}. However, despite theoretical proposals highlighting the potential of using material strain to mediate a large coupling between a two-level system and a mechanical degree of freedom \cite{Wilson-Rae}, an experimental demonstration of this strategy is still missing.

In this letter we present a novel hybrid system based on strain-mediated coupling. As shown in Fig.\ref{fig1}.a and detailed in the Methods section, it consists of a GaAs conical photonic nanowire embedding an optically active InAs quantum dot (QD). The QD can be considered as a two-level system with a ground state $|g\rangle$, and an excited state $|e\rangle$. As illustrated in Fig.\ref{fig1}.b and Fig.\ref{fig1}.c, the lowest order flexural vibration mode generates a time-varying strain field that modulates the QD transition energy $\hbar\omega$, thereby providing an effective strain-mediated coupling between the QD and the nanowire motion. This system is monolithic and compact, it requires no external parts to adjust the coupling strength between the quantum dot and the nanowire, and no external field to drive the coupling. Moreover, the nanowire conical geometry simultaneously confers a large light extraction efficiency with broadband operation \cite{Claudon}, and enhances the stress field in the vicinity of the clamped end, where the QD layer is situated. This hybrid system is well described by the independent spin-boson Hamiltonian \cite{Treutlein}:
\begin{equation}
H=\hbar\Omega_0(a^{\dagger}a+1/2)+\hbar\omega_0\sigma_z/2+\hbar g_0 \sigma_z (a^{\dagger}+a),
\end{equation}
where $\Omega_0/2\pi$ is the mechanical resonator eigenfrequency, $\hbar\omega_0$ the QD transition energy for the wire at rest, $a$ the phonon annihilation operator, $\sigma_z=|e\rangle\langle e| -|g\rangle\langle g|$ the Pauli operator of the two-level system. The last term describes the hybrid interaction, characterized by the hybrid vacuum coupling rate
\begin{equation}
g_0=\frac{\partial \omega}{\partial x}\Big\vert_{x=0}\delta x_{ZPF},
\label{eq1}
\end{equation}
where $x$ is the nanowire position, $x=0$ the rest position, $\hbar\omega$ the $x$-dependent QD transition energy, and $\delta x_{ZPF}$ the zero point fluctuations. This coupling results from the interaction between the excitonic transition and the lattice uniaxial deformation, which can be separated into two contributions \cite{Jons}: the hydrostatic component of the strain in the semiconductor lattice changes the QD bandgap, while the anisotropic components alter the valence band energy levels \cite{Bryant} (cf. Supplementary materials for details). Note that this system is effectively one-dimensional: due to a slight ellipticity of the nanowire cross-section, the lowest frequency vibration mode is split into a doublet of linearly polarized states. In this study we focus on the high frequency state which is polarized along $x$.

\begin{figure}[t]
\includegraphics[width=\columnwidth]{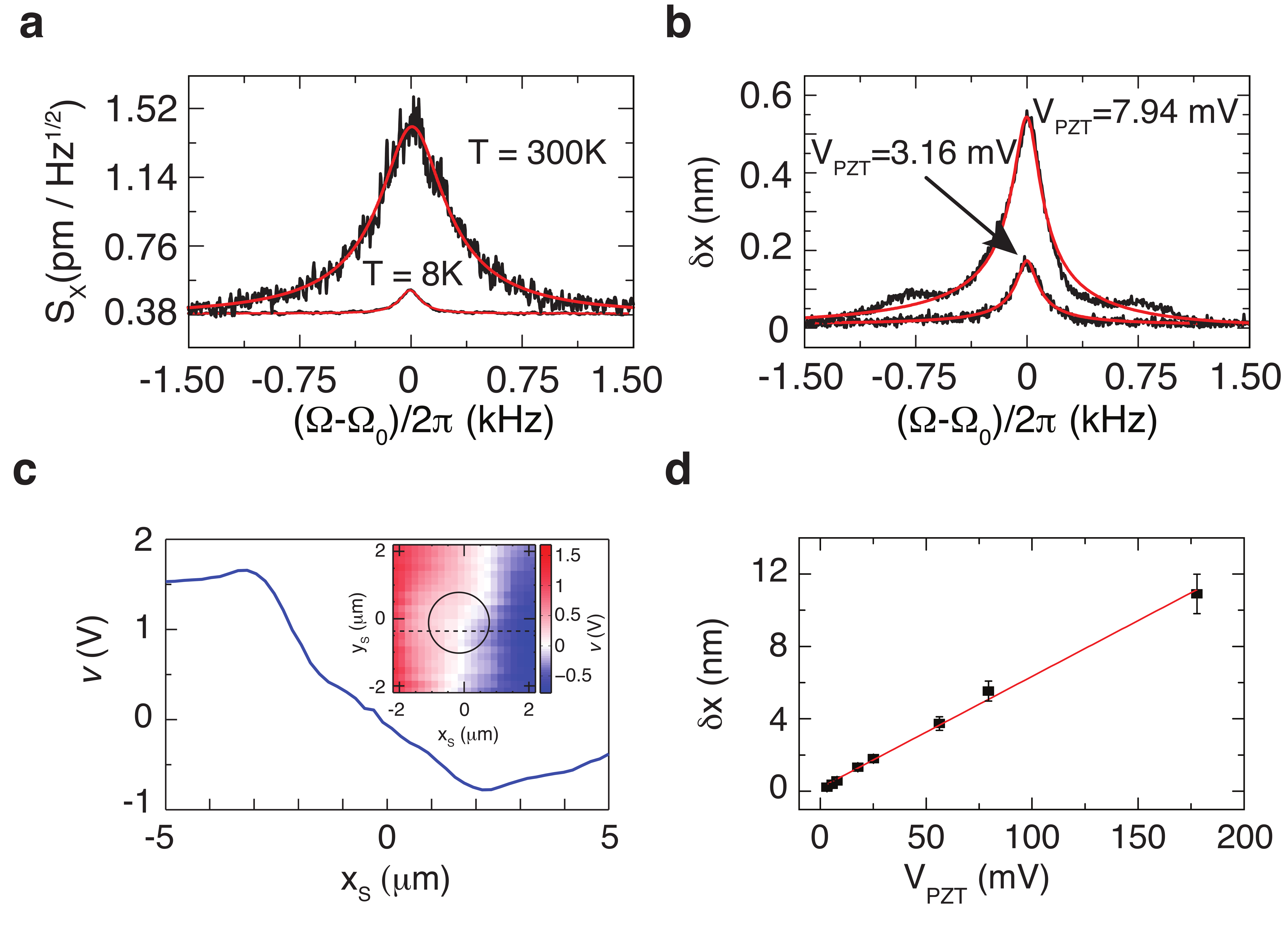}
\caption{\textbf{Nanowire mechanical properties}. \textbf{a} Displacement noise spectrum of the nanowire free-end $S_{x,th}[\Omega]$ measured in the Brownian motion regime, both at $T=8$K and $T=300$K. \textbf{b} Oscillation amplitude $\delta x$ of the nanowire free-end versus PZT drive frequency $\Omega/2\pi$ for two different PZT excitation amplitudes $V_{PZT}=3.16$meV and $V_{PZT}=7.94$meV ($T=8$K). The red lines are Lorentzian fits. The central frequency of the mode is $\Omega_0/2\pi=530\pm 0.5$kHz. \textbf{c} Static calibration of the displacement $\delta x$: the SPD voltage $v$ is plotted versus the laser spot position $x_s$ along a diameter of the wire top facet. In inset is shown the whole static calibration map $v(x_s,y_s)$. The circle indicates the top facet perimeter and the dashed line corresponds to the scanning path followed in the main panel. \textbf{d} Nanowire free-end displacement amplitude $\delta x$ versus PZT drive amplitude $V_{PZT}$ measured at $T=8$K.}
\label{fig2}
\end{figure}

In our experiment, a single nanowire is optically addressed (cf.Fig.\ref{fig1}.d and the Methods section). A probe laser beam is sent onto the free-end of the nanowire, enabling sensitive detection of its Brownian motion down to $T=8$K (Fig.\ref{fig2}.a). A piezoelectric transducer (PZT) can also be used to drive the nanowire oscillation in order to get its mechanical response spectrum (Fig.\ref{fig2}.b). We find that the eigenfrequency of the $x$-polarized mode is $\Omega_0/2\pi=530$kHz with a quality factor $Q=3000$ at $T=5$K, while at T=300K, due to temperature dependence of the mechanical properties and of phonon damping \cite{Mohanty}, the eigenfrequency downshifts to $\Omega_0/2\pi=522.4$kHz and $Q=1000$.


The evaluation of $g_0$ requires first a careful calibration of the motion amplitude $\delta x$ versus phonon number $\mathcal{N}=\delta x^2/2\delta x_{ZPF}^2$. To do so, we first calibrate the nanowire displacement $\delta x(V_{PZT})$ (Fig.\ref{fig2}.b and Fig.\ref{fig2}.d) using the static split-photodiode (SPD) voltage map shown in Fig.\ref{fig2}.c. This calibration allows us to precisely determine the conversion factor $\partial x/\partial V_{PZT}=(6.3\pm 0.5)\times 10^{-8}$m.V$^{-1}$. In a second step, we determine the thermomechanical variance of the displacement $\delta x_{th}^2$ by integrating the measured Brownian motion spectrum, $\delta x_{th}^2=\int S^2_{x,th}[\Omega,T]\mathrm{d}\Omega/2\pi$. The thermal average phonon number occupying a single mechanical mode at temperature $T$ is simply given by $\mathcal{N}(T)\simeq k_bT/\hbar\Omega_0$, yielding:
\begin{equation}
\delta x_{ZPF}=\delta x_{th}(T)\sqrt{\frac{\hbar\Omega_0}{2k_bT}},
\label{zpf}
\end{equation}
where $k_b$ is the Boltzmann constant. This procedure is applied at ambient and cryogenic temperatures ($T=300 K$ and $T=8 K$, see Fig.\ref{fig2}.a), and leads in both cases to $\delta x_{ZPF}=(1.1 \pm 0.3)\times 10^{-14}$m. This result is in excellent agreement with a finite element computation (see Methods section) which yields $\delta x_{ZPF}=1.2\times 10^{-14}$m.

\begin{figure}[t]
\includegraphics[width=\columnwidth]{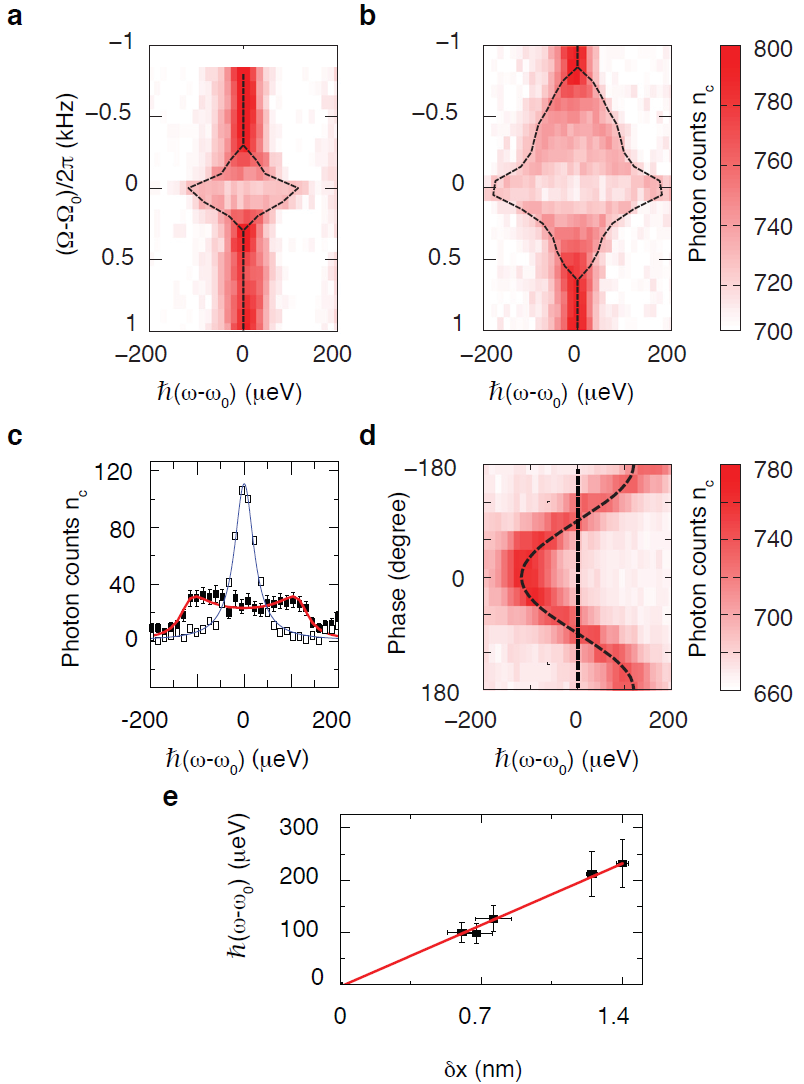}
\caption{\textbf{Characterization of the hybrid coupling}. Color coded photoluminescence spectra of a single quantum dot $n_c(\hbar\omega)$ versus mechanical excitation frequency $\Omega/2\pi$, \textbf{a} for maximum oscillation amplitude at resonance $\delta x_{max}=0.65$nm and \textbf{b} $\delta x_{max}=1.13$nm. $\hbar\omega_0=1390.25$meV is the quantum dot transition energy at rest. \textbf{c} Quantum dot spectra off the mechanical resonance, equal to the spectrum in absence of mechanical drive (hollow squares) and at resonance (filled squares). The solid lines are fits using eq.(\ref{camel}). \textbf{d} Stroboscopic measurement of the quantum dot energy shift versus the PZT drive modulation phase $\phi$. \textbf{e} QD spectral oscillation amplitude $\hbar\delta\omega$ versus nanowire free-end displacement amplitude $\delta x$.}
\label{fig3}
\end{figure}

We now examine the effect of the nanowire motion on a single quantum dot excitonic transition energy by micro-photoluminescence spectroscopy. Without mechanical excitation, the spectrum of each quantum dot is a Lorentzian peak of typical full width at half maximum $\hbar\Gamma=65\mu$eV (see Fig.\ref{fig3}.c). When the nanowire is driven by the PZT, we record series of photoluminescence spectra while sweeping the PZT frequency $\Omega$ across the mechanical resonance $\Omega_0$. The results are shown in Fig.\ref{fig3}.a and Fig.\ref{fig3}.b for two different oscillation amplitudes. At resonance ($\Omega=\Omega_0$), the quantum dot photoluminescence spectrum gets maximally broadened and deformed as a result of the time integrated oscillatory motion of the excitonic energy
\begin{equation}
n_c(\omega,\Omega)=i_0\int_0^\tau \mathcal{L}_\Gamma\left(\omega-\omega_0-\delta\omega(\Omega)\cos(\Omega t)\right)dt
\label{camel}
\end{equation}
where $n_c$ is the photon counts, $\tau\gg 2\pi/\Omega$ the integration time, $i_0$ the QD emission intensity, $\mathcal{L}_\Gamma(\omega)$ the Lorentzian function of linewidth $\hbar\Gamma$, $\hbar\omega_0$ the QD energy at rest, and $\hbar\delta\omega$ the oscillation amplitude of the quantum dot energy caused by the nanowire vibrations. As shown in Fig.\ref{fig3}.c this characteristic lineshape matches our measured spectra and allows accurate determination of the QD energy shift amplitude $\hbar\delta\omega(\Omega)$. This interpretation is confirmed by a stroboscopic measurement of the QD photoluminescence shown in Fig.\ref{fig3}.d (see Methods section for the experimental technique).

In order to extract the hybrid vacuum coupling rate $g_0$, we measure the maximal shift amplitude $\delta\omega_{max}=\delta\omega(\Omega_0)$ as a function of the nanowire oscillation amplitude $\delta x$ (Fig.\ref{fig3}.e). A linear behavior is obtained as expected in the low mechanical excitation regime. As illustrated in Fig.\ref{fig3}.b, the mechanical resonance spectral lineshape begins to turn asymmetric for larger excitation. At even higher excitation $\delta\omega_{max}(V_{PZT})$ saturates (not shown). These nonlinearities are likely the result of the dissipation induced heating of the wire: large motion amplitude causes a proportionally large temperature increase, which in turn shifts $\Omega_0$ away from the excitation frequency \cite{Arcizet_bistability}.


Using eq.(\ref{zpf}), we can rewrite eq.(\ref{eq1}) for $g_0$ in terms of the above measured quantities. Note that due to the inhomogeneous stress field in the quantum dot layer, $g_0$ depends drastically on the QD position within the nanowire: the closer to the sidewalls, the larger $g_0$. For the QD we studied in this work, we find a coupling strength $g_0/2\pi=450 \pm 100$kHz, which is nearly as large as the mechanical resonator frequency $\Omega_0/2\pi=530$kHz. This regime is very close to that of ultra-strong coupling \cite{Armour}, which is of importance for hybrid nanomechanical systems: the QD state dependent rest positions $x_e$ and $x_g$ of the nanowire are indeed spatially split by $|x_e-x_g|/\delta x_{ZPF}=g_0/\Omega_0$. In the ultra-strong coupling regime, $|x_e-x_g|$ exceeds the mechanical resonator wavefunction spreading $\delta x_{ZPF}$ in its ground state. This suggests, as in the case of a Stern-Gerlach experiment, that the QD state could be read out in a quantum non demolishing way through a sensitive position measurement \cite{Auffeves}. Another remarkable consequence of this large coupling strength is that the nanowire Brownian motion at T=5K may have a non-negligible effect on the QD spectral properties: according to our results, it contributes to the linewidth by $1\: \mu$eV, which is comparable to the radiative contribution.

So far, only a hybrid system based on electrostatic interaction and electrical readout \cite{Bennett} has shown a larger interaction strength. As detailed in the Supplementary information, future realistic improvements in the conical nanowire design can increase the coupling strength by a factor of 60, and the ultra strong coupling criterion $g_0/\Omega_0$ by a factor of 6. Indeed the stress experienced by the QD can be increased by (i) placing the QD layer exactly at the waist of the structure where the stress is maximum; (ii) decreasing the nanowire volume and lowering its aspect ratio towards shorter length and larger diameter; (iii) increasing the nanowire cone angle to stiffen further the upper part.

In conclusion, a semiconductor conical nanowire embedding QDs constitutes a novel kind of hybrid system where the coupling is mediated by the material strain. In addition to the original proposal by Wilson-Rae and coworkers \cite{Wilson-Rae}, Our approach has several important advantages. It combines a very high level of integration, full optical interface, and cryogenic compatibility. The conical shape enhances the hybrid coupling strength and offers as well a very efficient optical coupling between the QD and the outside world \cite{Claudon}. This is a favorable situation to envision quantum mechanical control of the nanowire motion using a single laser-driven QD \cite{Wilson-Rae}. Thanks to recent progress in diamond nanoprocessing, this approach could also be applied to NV-color-centers embedded in a diamond resonator \cite{Babinec} with the advantage that strain-mediated coupling potentially leads to a much larger coupling strength \cite{Batalov} than that based on magnetic field gradient used so far \cite{Rabl,Arcizet,Kolkowitz}. Thanks to the long spin lifetime in this sytem, such a coupling mechanism opens up the possibility of quantum non demolition readout of a single spin at room temperature \cite{Auffeves}. It also constitutes the basis for novel advanced quantum information processing schemes \cite{Stanigel}. Strain-mediated coupling could be implemented with profit in other mechanical micro or nanoresonator recently developed in the context of optomechanics, such as suspended photonic and phononic crystals \cite{Eichenfeld,Gavartin,Gavartin2}, microtoroids \cite{Schliesser} and microdisks \cite{Ding}.

\section{Methods}

\subsection{Conical Nanowire geometry}
The hybrid system studied in this work is made of epitaxial GaAs and has the shape of an inverted cone lying on a pyramidal pedestal. The inverted cone is $17.2\: \mu$m high, the diameter at the waist is $0.5\: \mu$m and the top facet diameter is $1.9\: \mu$m. This mechanical resonator embeds a single layer of self-assembled InAs QDs, located $0.8\: \mu$m above the waist (position determined by cathodoluminescence). From the quantum dot density, we estimate that the wire embeds approximately 100 QDs. To optimize the light extraction efficiency, the top facet features an anti-reflection coating ($115\: nm$ thick Si$_3$N$_4$ layer). To suppress spurious surface effects, the wire sidewalls are passivated and covered with a $20\: $nm thick Si$_3$N$_4$ layer \cite{Yeo}. We define such structures with a top-down process, very similar to the one described in \cite{Claudon}.

\subsection{Experimental set-up and calibration procedure}
The sample is located in vacuum on a cold finger at a temperature of $T=5$K. It can be mechanically excited by a piezo electrical transducer (PZT) glued at the back of the sample holder. The setup features two excitation/detection paths. The first one is dedicated to the mechanical motion spectroscopy, and the second one to the optical spectroscopy of single quantum dots. The different techniques employed are described below.

\textit{Mechanical motion spectroscopy} The free-end facet of the chosen nanowire is illuminated by a diffraction limited Gaussian spot using a continuous wave laser diode focused by a microscope objective of numerical aperture N.A.=0.4. The laser energy of 1.32eV (940nm) is chosen below the GaAs bandgap and the InAs wetting layer at 5K to prevent heating of the structure. The laser light is weakly reflected at the free-end facet interface due to index contrast with air. The reflection is sent onto a balanced split photodiode (SPD) with equal intensity on both quadrants. A voltage difference $v$ thus appears between both pixels such that, for a small displacement $\delta x$, $v \propto \delta x$. Therefore, using a spectrum and vectorial network analyzers, we can record both the displacement noise and the PZT driven response of the nanowire.

\textit{Position calibration procedure}
For the nanowire at rest, a static map of the SPD voltage difference $v$ versus the laser spot position $(x_s,y_s)$ is recorded using a piezo-controlled translation stage and an automatized scanning program. When the nanowire is in motion along $x$, a voltage difference $v_m(t)$ is measured; using the static map, this voltage difference can be related with the displacement $\delta x$ since $\delta x(t)=(\partial v/\partial x_s)^{-1}v_m(t)$.

\textit{Experimental determination of the zero point motion fluctuations }
The measurement of the Brownian motion spectrum provides us with a convenient way to determine the zero point motion fluctuation $\delta x_{ZPF}$ since the number of phonons in a thermal state $\mathcal{N}(T)$ is known from Bose statistics. For $kT\gg\hbar\Omega$ the latter simplifies into $\mathcal{N}\simeq kT/\hbar\Omega$, while the average energy of the zero point fluctuation amounts to $\hbar\Omega/2$. Hence, the zero point fluctuation can be derived from the Brownian motion according to the relation $\delta x_{th}(T)/\delta x_{ZPF}=\sqrt{2k_bT/\hbar\Omega}$, where $\delta x_{th}(T)$ is the root mean square displacement fluctuation of the Brownian motion.

For comparison, the theoretical value of $\delta x_{ZPF}$ is derived using a finite element commercial software. With this software, we evaluate the total elastic energy $\epsilon_{el}$ for a small known deflection $\delta x$ of the nanowire free-end. The zero point fluctuation $\delta x_{ZPF}$ is the standard deviation of the deflection when the average elastic energy amounts to $<\epsilon_{el}>=\frac{1}{2}\hbar\Omega$, such that
\begin{equation*}
\frac{1}{2}\hbar\Omega_0=\frac{\partial\epsilon_{el}}{\partial x^2}2\delta x^2_{ZPF}
\end{equation*}

\textit{Single quantum dot spectroscopy}
The photoluminescence of single quantum dot is excited with another continuous wave laser diode at 1.503 eV (825 nm), that creates free carriers in the InAs QDs wetting layer. Spectra (around 1.35 eV) are collected via the same objective and acquired on a charge coupled device (CCD) at the output of a 1.5 m focal length spectrometer operating with a 1200 grooves/mm grating. The overall spectral resolution amounts to 12 $\mu$eV which is in general well below the single QD linewidth.

\textit{Single quantum dot stroboscopy}
To acquire stroboscopic spectra of a single QD under mechanical excitation, the same excitation signal is used as a reference to drive the PZT and to modulate the excitation laser with an acousto-optic modulator. The "ON" time duration of the modulated excitation laser amounts to one tenth of the oscillation period $2\pi/\Omega_0$, effectively optically freezing the nanowire motion. Thus by changing the relative phase between the modulation signal and the mechanical drive signal, the QD spectrum can be acquired for any position $x$ of the nanowire free-end during an oscillation period.

\end{document}